# Two-dimensional ferroelastic semiconductors in $Nb_2SiTe_4$ and $Nb_2GeTe_4$ with promising electronic properties

Ting Zhang, Yandong Ma,* Xilong Xu, Chengan Lei, Baibiao Huang, and Ying Dai*

School of Physics, State Key Laboratory of Crystal Materials, Shandong University, Shandanan Street 27, Jinan 250100, China

Two-dimensional crystals with coupling of ferroelasticity and attractive electronic properties offer unprecedent opportunities for achieving long-sought controllable devices. But so far, the reported proposals are mainly based on hypothetical structures. Here, using first-principles calculations, we identify single-layer $Nb_2ATe_4$ (A = Si, Ge), which could be exfoliated from their layered bulks, are promising candidates. Single-layer $Nb_2ATe_4$ are found to be dynamically, thermally and chemically stable. They possess excellent ferroelasticity with high reversible ferroelastic strain and moderate ferroelastic transition energy barrier, beneficial for practical applications. Meanwhile, they harbor outstanding anisotropic electronic properties, including anisotropic carrier mobility and optical properties. More importantly, the anisotropic properties of single-layer $Nb_2ATe_4$ can be efficiently controlled through ferroelastic switching. These appealing properties combined with the experimental feasibility render single-layer $Nb_2ATe_4$ extraordinary platforms for realizing controllable devices.

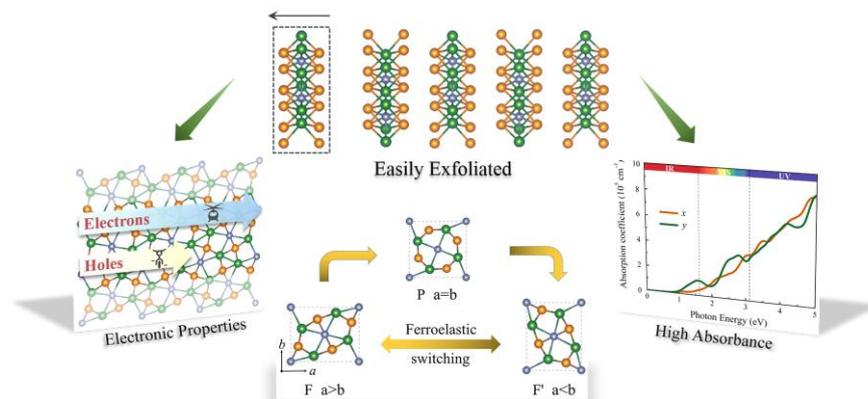

Two-dimensional (2D) ferroelastic crystals have attracted great interest for both fundamental researches and practical applications in nonvolatile memory devices.[1] The unique fingerprint of ferroelasticity is the existence of two or more equally stable orientation variants that can be interchanged without diffusion by applying stress.[2,3] Up to now, several 2D ferroelastic materials have been proposed, including t-YN, BP$_5$, GaTeCl and phosphorene.[4-7] Despite these efforts, currently, few 2D ferroelastic crystals have ever been verified in experiment. That is because most of the previous proposed systems either subject to air instability or are based on hypothetical structures.[6,8-10] For example, phosphorene, which should be a promising 2D ferroelastic crystal, suffers from rapid degradation under ambient conditions, thus challenging its practical applications.[8,9] Therefore, it great necessary to search for new 2D ferroelastic crystals, especially in the experimentally synthesized layered systems.

The recent interest in 2D ferroelastic crystals not only lies in ferroelasticity itself but also extends to the coupling of ferroelasticity and electronic properties.[11-14] Compared with 2D crystals presenting ferroelastic order only, undoubtedly, 2D crystals that hold ferroelasticity and intriguing electronic properties simultaneously are more desirable as they will open up unprecedented opportunities for developing multifunctional and controllable devices. Currently, there are mainly two types of such coupling in 2D crystals. One is the coupling between ferroelasticity and other ferroic orders (i.e., ferroelectricity and ferromagnetism), namely multiferroicity.[15,16] 2D multiferroic crystals hold potential for applications in nonvolatile memory devices because they could overcome the quantum tunneling effect and power dissipation encountered in conventional memory devices.[17] Examples mainly include BP$_5$, Janus VSSe, silver and copper monohalides.[7,12,17] Another is the coupling between ferroelasticity and nontrivial topology.[18] In our previous work, we proposed that single-layer 1S'-MSSe (M = Mo, W) is a 2D ferroelastic topological insulator and holds appealing potential for controlling the anisotropy of the topological edge state via ferroelastic switching.[18] The few existing types of such coupling, combined with the fact that most of these proposals are based on hypothetical structures, severely limits their further developments and applications.

In this work, on the basis of first-principles calculations, we report the identification of a novel family of 2D ferroelastic crystals in single-layer (SL) Nb$_2$ATe$_4$ (A = Si, Ge). The ferroelasticity in SL Nb$_2$SiTe$_4$ (Nb$_2$GeTe$_4$) shows a significant ferroelastic signal with a reversible strain as high as 24.4% (22.1%) and a moderate ferroelastic switching barrier of 0.237 (0.173) eV/atom, suggesting that they are excellent 2D ferroelastic crystals. In addition to ferroelasticity, SL Nb$_2$ATe$_4$ is found to be an indirect-gap semiconductor with intriguing anisotropic electronic properties, i.e., anisotropic carrier mobility and optical properties. Such coexistence of ferroelasticity and anisotropic electronic properties in SL Nb$_2$ATe$_4$ facilitates the coupling between them, namely, ferroelastic switching can preciously control its anisotropic electronic properties, which are useful for designing controllable device. Moreover, SL Nb$_2$ATe$_4$ exhibits high experimental feasibility as it can be exfoliated from its layered bulk.

All calculations are performed on basis of density functional theory (DFT) implemented in Vienna ab initio simulation package (VASP).[19] The exchange and correlation functional is described within the generalized gradient approximation (GGA) in form of Perdew Burke Ernzerhof functional (PBE).[20,21] A vacuum space of 20 Å is adopted to minimize the interlayer interaction between adjacent layers. The van der Waals (vDW) interactions are corrected by the DFT-D2 approach.[22] The plane-wave cut-off energy is set as 500 eV. The convergence criteria for energy and force are set to $10^{-5}$ eV and 0.01eV/Å, respectively. A 4 × 5 × 1 k-point grid in the Monkhorst-Pack scheme is used to sample the Brillouin zone for structure optimizations. And a denser k-point grid of 7 × 9 × 1 is adopted for electronic structure calculations. Due to PBE functional usually underestimates the band gap, we study the band structures and optical properties based on the Heyd Scuseria Ernzerhof (HSE06) hybrid functional.[23] AIMD simulations are performed at 500 K by using a NVT ensemble that last 5 ps with a time step of 2 fs.[24] The phonon spectra is calculated by the PHONOPY code.[25]

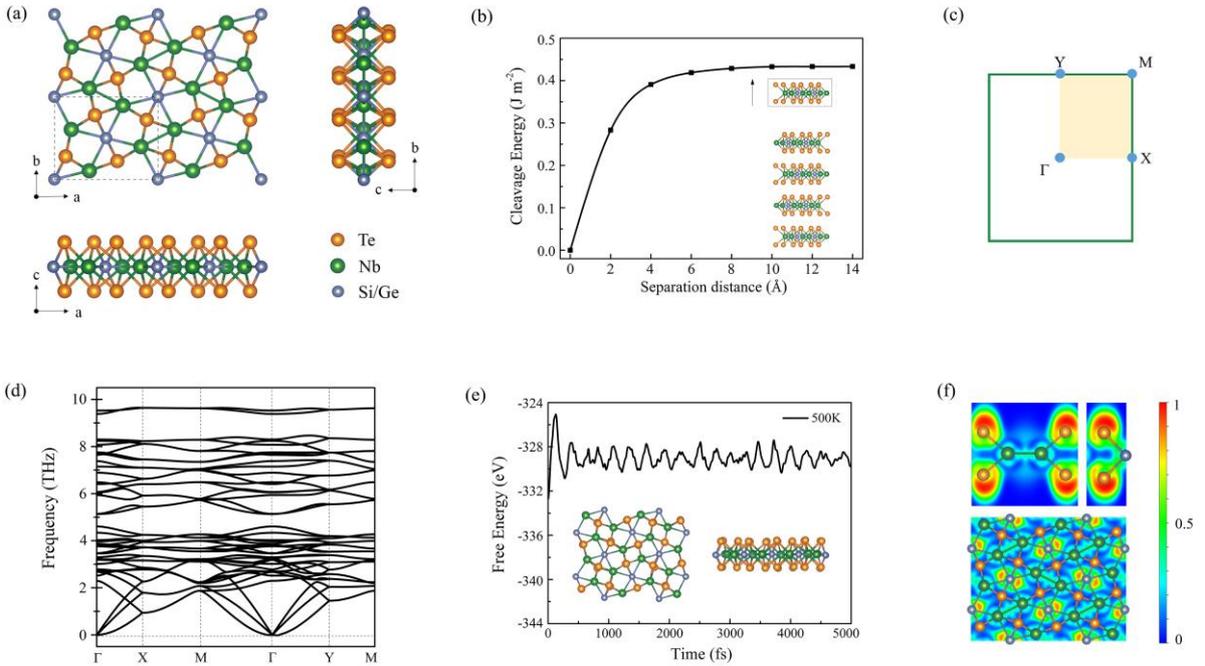

**Fig. 1** (a) Crystal structure of SL Nb$_2$ATe$_4$ from top and side views. (b) The cleavage energy $E_{cl}$ varies with separation distance in the process of exfoliating single-layer Nb$_2$SiTe$_4$. Inset in (b) is the schematic of exfoliating single layer from the five-layer system. (c) 2D Brillouin zone. (d) Phonon spectra of SL Nb$_2$SiTe$_4$. (e) Fluctuation of total energy with time in the AIMD simulation of SL Nb$_2$SiTe$_4$ at 500 K and its corresponding snapshot taking from the end of the simulation. (e)

Two-dimensional views of the electron localization function of SL Nb$_2$SiTe$_4$.

Bulk Nb$_2$ATe$_4$ was synthesized experimentally in 1990s.[26,27] The crystal structure of bulk Nb$_2$ATe$_4$ is shown in **Fig. S1(a)**. It crystallizes in monoclinic structure with space group $P2_1/c$, forming a layered configuration. Nb$_2$ATe$_4$ layers are stacked via vdW interaction along the c axis. Within Nb$_2$ATe$_4$ layer, the planar sublayer composed of Nb and A atoms is sandwiched between two Te atomic sublayers. Each A atom is coordinated with four Nb atoms and eight Te atoms. **Fig. S1** shows the band structures of bulk Nb$_2$SiTe$_4$ (Nb$_2$GeTe$_4$), which are semiconductors with a band gap of 0.34 (0.24) eV based on PBE functional. Considering the strong quasi-2D anisotropy of bulk Nb$_2$ATe$_4$, its SL structure might be obtained easily. To evaluate the experimental feasibility of exfoliating the SL structure, we calculated the cleavage energy ($E_{cl}$) by imposing a fracture in a five-layer slab of Nb$_2$ATe$_4$; see **Fig. 1(b)** and **S2(a)**. The obtained $E_{cl}$ of SL Nb$_2$SiTe$_4$ (Nb$_2$GeTe$_4$) is 0.43 (0.42) J/m$^2$, quite close to that of graphene (0.37 J/m$^2$),[28] MoS$_2$ (0.42 J/m$^2$)[29] and GeS (0.52 J/m$^2$)[30]. This firmly indicates that Nb$_2$ATe$_4$ shows a rather weak interlayer interaction and the isolation of SL Nb$_2$ATe$_4$ is experimentally feasible via mechanical or liquid exfoliation.[31]

**Fig. 1(a)** presents the crystal structure of SL Nb$_2$ATe$_4$. The optimized lattice constants of SL Nb$_2$SiTe$_4$ (Nb$_2$GeTe$_4$) are found to be a = 7.96 (8.00) Å and b = 6.40 (6.55) Å. To assess the dynamical stability of SL Nb$_2$ATe$_4$, we calculate its phonon spectra, which are shown in **Fig. 1(d)** and **S2(b)**. The phonon spectrums contain no imaginary frequency, indicating that SL Nb$_2$ATe$_4$ is dynamically stable. We also perform AIMD simulations to check the thermal stability of SL Nb$_2$ATe$_4$. After heating at 500 K for 5 ps with a time step of 2 fs, the structures show neither bond broken nor significant structural transformation [see **Fig. 1(e)** and **S2(c)**], suggesting the thermal stability of SL Nb$_2$ATe$_4$. To gain deep insight into its stability at ambient conditions, we further carry out AIMD simulations for SL Nb$_2$ATe$_4$ exposed to gas-phase O$_2$ at 300 K. Here, 18 O$_2$ molecules (within a 2 × 2 × 1 supercell) are initially placed about 4 Å above the surfaces of SL Nb$_2$ATe$_4$. As shown in **Fig. S3**, after the simulation, the O$_2$ molecules move away from the surfaces without spontaneous dissociation into oxygen atoms, which implies good stability of SL Nb$_2$ATe$_4$ against oxidization. And we also plot electron localization function (ELF) maps in the horizontal plane, Nb-Te plane and A-Te plane to identify the bonding features. As shown in **Fig. 1(f)** and **S2(d)**, the electron localizations are mainly distributed around the Te and A atoms, suggesting ionic characters for the Te-Nb and A-Nb bonds, and covalent characters for A-Te bonds.

The mechanical properties of SL Nb$_2$ATe$_4$ are then investigated. The calculated elastic constants of SL Nb$_2$SiTe$_4$ (Nb$_2$GeTe$_4$) are $C_{11}$ = 92.7 (86.9) N/m, $C_{22}$ = 96.5 (87.0) N/m, $C_{12}$ = 21.8 (21.0) N/m and $C_{66}$ = 45.9 (42.3) N/m, which meet the Born criteria: $C_{11}C_{22} - C_{12}^2 > 0$ and $C_{66} > 0$.[32,33]

Based on the elastic constants, the Young's modulus $Y(\theta)$ and Possion's ratio $\upsilon(\theta)$ of SL $Nb_2ATe_4$ along the in-plane $\theta$ can be obtained by the following expression,[34,35]

$$Y(\theta) = \frac{C_{11}C_{22} - C_{12}^2}{C_{11}\sin^4\theta + A\sin^2\theta\cos^2\theta + C_{22}\cos^4\theta},$$

$$\upsilon(\theta) = \frac{C_{12}\sin^4\theta - B\sin^2\theta\cos^2\theta + C_{12}\cos^4\theta}{C_{11}\sin^4\theta + A\sin^2\theta\cos^2\theta + C_{22}\cos^4\theta},$$

where $A = (C_{11}C_{22} - C_{12}^2)/C_{66} - 2C_{12}$ and $B = C_{11} + C_{22} - (C_{11}C_{22} - C_{12}^2)/C_{66}$. $\theta = 0°$ corresponds to the a axis. The angular dependent results of $Y(\theta)$ and $\upsilon(\theta)$ are presented in **Fig. 2(a, b)** and **S4**. We can see that the Young's modulus of SL $Nb_2SiTe_4$ ($Nb_2GeTe_4$) range from 87.7 (81.8) to 102.7 (94.9) N/m, showing a slight mechanical anisotropy. These values are smaller than those of graphene (340 N/m)[36] and BN (318 N/m)[37], which implies their mechanical flexibility and thus the feasibility of ferroelastic switching via applying external stress. While for the Poisson's ratio, it varies from 0.12 to 0.24, which is in the reasonable range (0~0.5).[38] This suggests their moderate structural response to external stress, also beneficial for ferroelasticity.

**Table 1** The lattice constants (*a*, *b*), ferroelastic strains and ferroelatic switching barriers of SL $Nb_2ATe_4$.

|  | *a* (Å) | *b* (Å) | Ferroelastic strain | Energy barrier (eV/atom) |
|---|---|---|---|---|
| $Nb_2SiTe_4$ | 7.96 | 6.40 | 24.4% | 0.237 |
| $Nb_2GeTe_4$ | 8.00 | 6.55 | 22.1% | 0.173 |

Having examined the stability and mechanical properties of SL $Nb_2ATe_4$, we investigate its ferroelastic properties. The ferroelasticity in SL $Nb_2ATe_4$ stems from the fact that its centrosymmetric state (S.G. *P4/mbm*) is not stable and would experience spontaneous structure transformation. This spontaneous transformation can occur along either a or b axis, resulting in two degenerate ferroelastic ground states (S.G. *Pbam*), F and F'; see **Fig. 2(c)**. For convenience of our discussion, here we take F as initial state and F' as final state. For initial state F, as listed in **Table 1**, the lattice constant a is larger than b. Upon applying external tensile stress along the b axis, it can transform into final state F' where |a'| = |b|, |b'| = |a|. Accordingly, the shorter axis is switched to the a axis, and its crystal structure is same as initial state F with rotating 90°. Similarly, the inverse structure transformation from final state F' to initial state F can also occur when applying uniaxial tensile stress along the a axis. The centrosymmetric state P with a square lattice can be considered as the paraelastic state for SL Nb$_2$SiTe$_4$ (Nb$_2$GeTe$_4$), whose lattice constants are found to be a = b = 7.14 (7.32) Å, as shown in **Fig. 2(c)**.

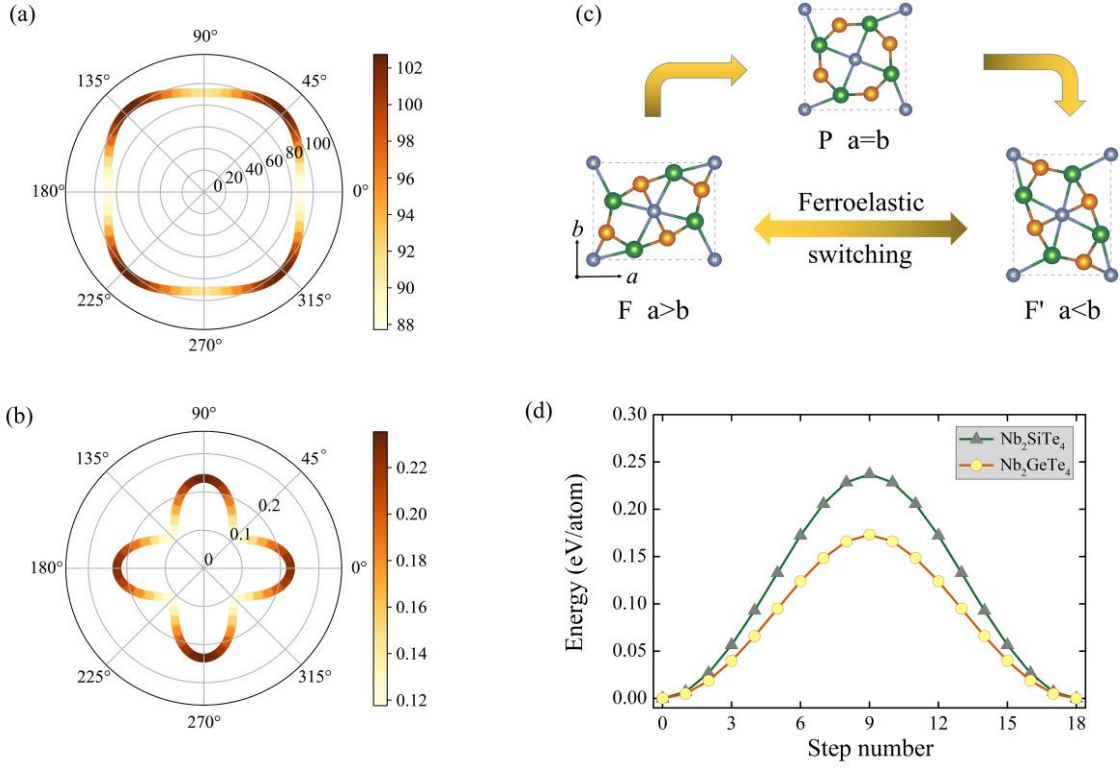

**Fig. 2** (a) Young's modulus and (b) Possion's ratio of SL Nb$_2$SiTe$_4$ vary with the in-plane angle $\theta$ ($\theta$ = 0° corresponds to the x axis). (c) The pathway of ferroelastic switching for SL Nb$_2$ATe$_4$. (d) The energy barriers of ferroelastic switching based on NEB method for SL Nb$_2$ATe$_4$.

To estimate the ferroelastic signal intensity of SL Nb$_2$ATe$_4$, we investigate its reversible ferroelastic strain, which is defined as [(|a|/|b|-1)×100%]. Apparently, a strong ferroelastic signal, that is a high reversible strain, will be beneficial for the ferroelastic performance. The ferroelastic reversible strain for SL Nb$_2$SiTe$_4$ (Nb$_2$GeTe$_4$) is found to be as high as 24.4% (22.1%), comparable with that of t-YN[5], GeS[6] and InOY (Y = Cl, Br)[39], implying the strong ferroelastic signal of SL Nb$_2$ATe$_4$. Another key factor that affects the ferroelastic performance is switching barrier, which determines the transformation processes for the lattice orientation switch. We thus study the energy barrier for ferroelastic switching from initial state F to final state F' of SL Nb$_2$ATe$_4$ [**Fig. 2(b)**] using the climbing image nudged elastic band (NEB) method.[40] As shown in **Fig. 2(d)**, in view of the structure symmetry, the pathway from P to F is identical to the path from P to F'. The energy barrier for SL Nb$_2$SiTe$_4$ (Nb$_2$GeTe$_4$) is calculated to be 0.237 (0.173) eV/atom, which is higher than those of t-YN (33 meV/atom)[5] and borophane (0.1 eV/atom)[41], but lower than that of BP$_5$ (0.32 eV/atom)[7] and comparable to that of phosphorene (0.20 eV/atom)[6]. This highlights the desirable possibility of ferroelastic switching under external stress and its robustness against environmental perturbation.

The ferroelastic switching in SL Nb$_2$ATe$_4$ can be further inspected from transformation strain matrix that uses the paraelastic structure as a reference state. The matrix can be obtained by the Green Lagrange strain tensor,[42]

$$\eta_x = \frac{1}{2}\left(\left[H_{ref}^{-1}\right]^T H_x^T H_x H_{ref}^{-1} - I\right),$$

where $I$ is a 2 × 2 identity matrix, and $H_x$ and $H_{ref}$ are lattice constants of the ferroelastic and paraelastic states, respectively. $H_x$ and $H_{ref}$ of SL Nb$_2$SiTe$_4$ (Nb$_2$GeTe$_4$) are [7.96, 0; 0, 6.40] ([8.00, 0; 0, 6.55]) and [7.14, 0; 0, 7.14] ([7.32, 0; 0, 7.32]), respectively. The corresponding strain matrix $\eta_x$ for initial state F of SL Nb$_2$SiTe$_4$ and Nb$_2$GeTe$_4$ can be described respectively as:

$$\eta_x = \begin{bmatrix} 0.121 & 0 \\ 0 & -0.098 \end{bmatrix} \text{ and } \eta_x = \begin{bmatrix} 0.097 & 0 \\ 0 & -0.100 \end{bmatrix}.$$

This indicates that there is a 12.1% (9.7%) tensile strain along the a axis and a 9.8% (10.0%) compressive strain along the b axis. Similarly, the strain matrix $\eta_y$ for final state F' of SL Nb$_2$SiTe$_4$ and Nb$_2$GeTe$_4$ is expressed respectively as follows:

$$\eta_y = \begin{bmatrix} -0.098 & 0 \\ 0 & 0.121 \end{bmatrix} \text{ and } \eta_y = \begin{bmatrix} -0.100 & 0 \\ 0 & 0.097 \end{bmatrix}.$$

And a 9.8% (10.0%) compressive strain along the a axis and a 12.1% (9.7%) tensile strain along the b axis are obtained. These values are significantly larger than many other 2D ferroelastic materials such as GeSe (2.7% and 4.1% transformation strains),[6] indicating the excellent ferroelastic performance of SL Nb$_2$ATe$_4$.

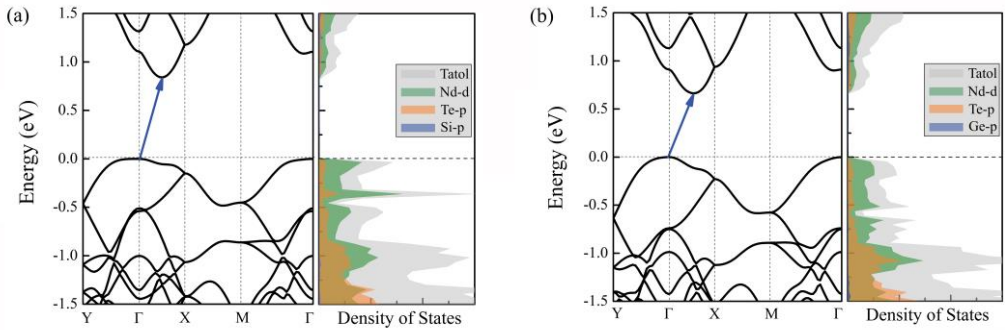

**Fig. 3** Band structures and projected density of states of (a) SL Nb$_2$SiTe$_4$ and (b) SL Nb$_2$GeTe$_4$. The Fermi level is set to zero.

We then investigate the electronic properties of SL Nb$_2$ATe$_4$. The band structures and projected density of states (PDOS) of SL Nb$_2$ATe$_4$ based on HSE06 functional are shown in **Fig. 3(a, b)**. SL

Nb$_2$SiTe$_4$ is an indirect gap semiconductor with a band gap of 0.84 eV. Its valence band maximum (VBM) locates at the Γ point, and conduction band minimum (CBM) locates between the Γ and X points. While for SL Nb$_2$GeTe$_4$, it also possesses an indirect band gap of 0.66 eV with VBM locating at the Γ point and CBM lying along the Γ-X path. On the basis of the projected density of states, we find that the VBM and CBM are mainly dominated by the p-orbital of Te atom and the d-orbital of Nb atom, respectively. Another interesting point of the band structures that should be noted is that the valence band edges of SL Nb$_2$ATe$_4$ are less dispersed with respected to that of its conduction band edges. It strongly implies the anisotropic electronic properties of SL Nb$_2$ATe$_4$. To that end, we investigate the carrier effective mass, which can be obtained by the following equation:

$$m^*_{e(h)} = \pm \hbar^2 (\frac{d^2 E_k}{dk^2})^{-1}.$$

Here $k$ is wave vector, and $E_k$ is the carrier energy corresponding to $k$. The corresponding results are summarized in **Table 2**. The effective masses of electrons are 0.47 m$_e$ (0.29 m$_e$) for SL Nb$_2$SiTe$_4$ and 0.66 m$_e$ (0.32 m$_e$) for SL Nb$_2$GeTe$_4$ along the x (y) axis. And the effective masses of holes are 1.30 m$_e$ (4.20 m$_e$) for SL Nb$_2$SiTe$_4$ and 0.77 m$_e$ (121.9 m$_e$) for SL Nb$_2$GeTe$_4$ along the x (y) axis. The anisotropic ratios of the effective masses of carriers in SL Nb$_2$SiTe$_4$ (1.62 and 3.23) and Nb$_2$GeTe$_4$ (2.06 and 158.3) are significant,[43] which indicates the anisotropic electronic behaviors.

**Table 2** In-plane stiffness ($C$), effective mass ($m^*$), deformation potential ($E_d$) and carrier mobility ($\mu$) of SL Nb$_2$ATe$_4$.

| System | Carrier type | Direction | $C$ (N m$^{-1}$) | $m^*$ | $E_d$ (eV) | $\mu$ (cm$^2$ V$^{-1}$ s$^{-1}$) |
|---|---|---|---|---|---|---|
| Nb$_2$SiTe$_4$ | hole | x | 92.67 | 1.30 | 2.85 | 79.30 |
| | | y | 97.06 | 4.20 | 0.95 | 2.35×10$^2$ |
| | electron | x | 92.67 | 0.47 | 1.58 | 4.57×10$^3$ |
| | | y | 97.06 | 0.29 | 2.16 | 4.04×10$^3$ |
| Nb$_2$GeTe$_4$ | hole | x | 86.90 | 0.77 | 2.59 | 37.32 |
| | | y | 87.15 | 121.9 | 0.73 | 2.98 |
| | electron | x | 86.90 | 0.66 | 1.80 | 1.91×10$^3$ |
| | | y | 87.15 | 0.32 | 2.40 | 2.21×10$^3$ |

To explore the electronic transport behaviors of SL Nb$_2$ATe$_4$, we estimate their carrier mobility based on the deformation potential (DP) theory:[44,45]

$$\mu = \frac{e\hbar^3 C}{k_B T m m_d E_d^2}.$$

Here, $k_B$ is the Boltzmann constant, $T$ is 300K, m is the effective mass of carriers in the transport direction, and $m_d = \sqrt{m m_\perp}$ is the carrier average effective mass. The in-plane stiffness is defined as $C = (\partial^2 E/\partial \varepsilon^2)/S_0$, where $E$ and $S_0$ are the total energy and the lattice area. The deformation potential, defined as $E_d = dE_{edge}/d\varepsilon$, denotes the shift of the band edge $E_{edge}$ with respect to uniaxial strain $\varepsilon$. As listed in **Table 2**, the electron mobility of SL Nb$_2$ATe$_4$ can reach as high as 1.91~4.57×10$^3$ cm$^2$ V$^{-1}$ s$^{-1}$, an order of magnitude larger than that of MoS$_2$ (320 cm$^2$ V$^{-1}$ s$^{-1}$)[46] and close to that of phosphorene (2.2 × 10$^3$ cm$^2$ V$^{-1}$ s$^{-1}$)[47]. The high carrier mobility makes SL Nb$_2$ATe$_4$ very promising for applications in 2D electronic devices. Different from the cases of electrons, the hole mobility shows significant anisotropy. The hole mobility of SL Nb$_2$SiTe$_4$ along y direction is an order of magnitude larger than that along x direction. While for SL Nb$_2$GeTe$_4$, the hole mobility along y direction is an order of magnitude smaller than that along x direction. Therefore, SL Nb$_2$ATe$_4$ exhibits ferroelasticity and anisotropic electronic properties simultaneously and it is expected to realize the precise direction-control of the hole transport using ferroelastic switching. Upon introducing ferroelastic switching, the lattice orientations of SL Nb$_2$ATe$_4$ are rotated by 90°. And the favorable transport direction for holes also experiences a 90° rotation. As a result, the transport direction of the holes in SL Nb$_2$ATe$_4$ is switchable between the x and y directions via ferroelastic switching, providing promising candidates for designing controllable electronic devices.

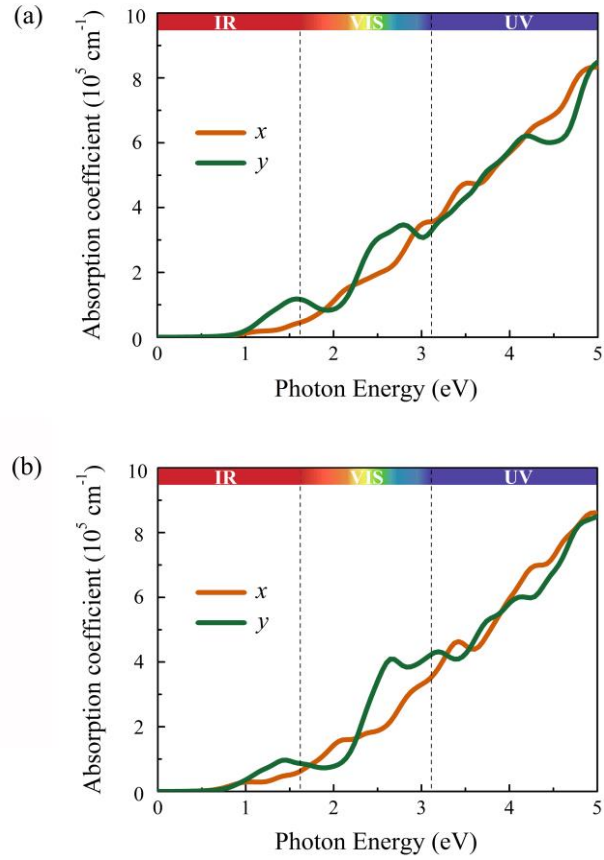

**Fig. 4** Optical absorption coefficients of (a) SL Nb$_2$SiTe$_4$ and (b) SL Nb$_2$GeTe$_4$.

Besides, considering the moderate band gap of SL Nb$_2$ATe$_4$, it would be interesting to investigate its optical properties. We then calculate the optical absorption coefficients of SL Nb$_2$ATe$_4$ using the following expression[48]

$$\alpha(\omega) = \sqrt{2}\frac{\omega}{c}[\sqrt{\varepsilon_1^2(\omega)+\varepsilon_2^2(\omega)}-\varepsilon_1(\omega)]^{1/2},$$

where $\omega$ refers to the frequency, and $\varepsilon_1(\omega)$ and $\varepsilon_2(\omega)$ are the real and imaginary parts in the frequency dependent dielectric function, respectively. As shown in **Fig. 4**, SL Nb$_2$ATe$_4$ exhibits a remarkably high optical absorbance from ultraviolet to infrared regions. The absorption coefficient values reach the order of ~$10^5$ cm$^{-1}$, implying excellent light harvesting ability. In addition, as shown in **Fig. 4**, the optical absorption of SL Nb$_2$ATe$_4$ shows strong anisotropy, namely, the optical absorption along y direction is much stronger than that along x direction. This can be attributed to its anisotropic morphology. The SL Nb$_2$ATe$_4$ thus can be used for applications in polarized optical sensors. Moreover, the coexistence of anisotropic optical behavior and ferroelasticity in SL Nb$_2$ATe$_4$ makes the coupling between them also possible. Like the case of anisotropic transport property, by introducing ferroelastic switching in SL Nb$_2$ATe$_4$, its anisotropic optical behavior would experience a 90° rotation. Therefore, the precise direction-control of optical behaviors using ferroelastic switching can be achieved in SL $Nb_2ATe_4$.

In summary, using first-principles calculations, we systematically investigate the ferroelastic, electronic and optical properties of SL $Nb_2ATe_4$. We demonstrate that SL $Nb_2ATe_4$ can be easily exfoliated from its bulk owing to its small exfoliation energy. And it is found to be dynamically, thermally and chemically stable. We reveal that SL $Nb_2SiTe_4$ ($Nb_2GeTe_4$) shows excellent ferroelasticity with a high reversible ferroelastic strain of 24.4% (22.1%) and a moderate ferroelastic transition energy barrier of 0.237 (0.173) eV/atom. We also find that SL $Nb_2SiTe_4$ ($Nb_2GeTe_4$) is an indirect-gap semiconductor with a band gap of 0.84 (0.66) eV and shows remarkable anisotropic electronic properties, including anisotropic carrier mobility and optical properties. We further show that these anisotropic properties of single-layer $Nb_2ATe_4$ can be precisely controlled through ferroelastic switching, achieving the coupling between them. Our findings provides an ideal platform for exploring ferroelasticity, anisotropic behavior and their coupling.

## ASSOCIATED CONTENT

Supporting Information Available: Relevant properties for $Nb_2ATe_4$ (A = Si, Ge) bulk and SL $Nb_2GeTe_4$.

## AUTHOR INFORMATION

Corresponding Authors

*E-mail: yandong.ma@sdu.edu.cn (Y.M.).

*E-mail: daiy60@sina.com (Y.D.).

ORCID

Yandong Ma: 0000-0003-1572-7766

Ying Dai: 0000-0002-8587-6874

Notes

The authors declare no competing financial interest.

# ACKNOWLEDGMENTS

This work is supported by the National Natural Science Foundation of China (No. 11804190), Shandong Provincial Natural Science Foundation of China (Nos. ZR2019QA011 and ZR2019MEM013), Qilu Young Scholar Program of Shandong University, and Taishan Scholar Program of Shandong Province, and Youth Science and Technology Talents Enrollment Project of Shandong Province.

# REFERENCES

(1) Wang, H.; Qian, X. Two-Dimensional Multiferroics in Monolayer Group IV Monochalcogenides. *2D Mater*. 2017, 4, 015042.
(2) Salje, E. Ferroelastic Materials. *Annu. Rev. Mater. Res.* 2012, 42, 265-283.
(3) Poquette, B.; Asare, T.; Schultz, J.; Brown, D.; Kampe, S. Domain Reorientation as a Damping Mechanism in Ferroelastic-Reinforced Metal Matrix Composites. *Metall. Mater. Tran. A* 2011, 42, 2833-2842.
(4) Zhang, S.-H.; Liu, B.-G. A controllable Robust Multiferroic GaTeCl Monolayer with Colossal 2D Ferroelectricity and Desirable Multifunctionality. *Nanoscale* 2018, 10, 5990-5996.
(5) Xu, B.; Xiang, H.; Yin, J.; Xia, Y.; Liu, Z. A Two-Dimensional Tetragonal Yttrium Nitride Monolayer: A Ferroelastic Semiconductor with Switchable Anisotropic Properties. *Nanoscale* 2018, 10, 215-221.
(6) Wu, M.; Zeng, X.; Intrinsic Ferroelasticity and/or Multiferroicity in Two-Dimensional Phosphorene and Phosphorene Analogues. *Nano Lett.* 2016, 16, 3236-3241.
(7) Wang, H.; Li, X.; Sun, J.; Liu, Z.; Yang, J. $BP_5$ Monolayer with Multiferroicity and Negative Poisson's Ratio: A Prediction by Global Optimization Method. *2D Mater.* 2017, 4, 045020.
(8) Koenig, S. P.; Doganov, R. A.; Schmidt, H.; Castro Neto, A. H.; Özyilmaz, B., Electric Field Effect in Ultrathin Black Phosphorus. *Appl. Phys. Lett.* 2014, 104, 103106.
(9) Ziletti, A.; Carvalho, A.; Campbell, D. K.; Coker, D. F.; Castro Neto, A. H., Oxygen Defects in Phosphorene. *Phys. Rev. Lett.* 2015, 114, 046801.
(10) Zhang, T.; Ma, Y.; Huang, B.; Dai, Y. Two-Dimensional Penta-$BN_2$ with High Specific Capacity for Li-Ion Batteries. *ACS Appl. Mater. Inter.* 2019, 11, 66104-6110.
(11) Zhang, Y.; Lu, H.; Xie, L.; Yan, X.; Paudel, T.R.; Kim, J.; Cheng, X.; Wang, H.; Heikes, C.; Li, L.; Xu, M.; Schlom, D. G.; Chen, L.-Q.; Wu, R.; Tsymbal, E. Y.; Gruverman, A.; Pan, X. Anisotropic Polarization-Induced Conductance at a Ferroelectric-Insulator Interface. *Nat. Nanotechnol.* 2018, 13, 1132-1136.
(12) Zhang, C.; Nie, Y.; Sanvito, S.; Du, A. First-Principles Prediction of a Room-Temperature Ferromagnetic Janus VSSe Monolayer with Piezoelectricity, Ferroelasticity, and Large Valley Polarization. *Nano Lett.* 2019, 19, 1366-1370
(13) Zhao, Y.; Zhang, J.-J.; Yuan, S.; Chen, Z. Nonvolatile Electrical Control and Heterointerface-Induced Half-Metallicity of 2D Ferromagnets. *Adv. Funct. Mater.* 2019, 29, 1901420.
(14) Zhang, Z.; Ni, X.; Huang, H.; Hu, L.; Liu, F. Valley Splitting in the Van der Waals Heterostructure $WSe_2/CrI_3$: The Role of Atom Superposition. *Phys. Rev. B* 2019, 99, 115441.
(15) Wang, K.; Liu, J.; Ren, Z. Multiferroicity: the Coupling between Magnetic and Polarization Orders. *Adv. Phys.* 2009, 58, 321-448.
(16) Yamauchi, K.; Picozzi, S. Interplay between Charge Order, Ferroelectricity, and Ferroelasticity: Tungsten Bronze Structures as a Playground for Multiferroicity. *Phys. Rev. Lett.* 2010, 105, 107202.
(17) Gao, Y.; Wu, M.; Zeng, X.-C. Phase Transitions and Ferroelasticity-Multiferroicity in Bulk and Two-Dimensional Silver and Copper Monohalides. *Nanoscale Horiz.* 2019, 4, 1106-1112.

(18) Ma, Y.; Kou, L.; Huang, B.; Dai, Y.; Heine, T. Two-dimensional Ferroelastic Topological Insulators in Single-Layer Janus Transition Metal Dichalcogenides MSSe (M = Mo,W). *Phys. Rev. B* 2018, 98, 085420.

(19) Kresse, G.; Furthmüller, J. Efficient Iterative Schemes for Ab Initio Total-Energy Calculations using a Plane-Wave Basis Set. *Phys. Rev. B* 1996, 54, 11169-11186.

(20) Perdew, J.; Burke, K.; Ernzerhof, M. Generalized Gradient Approximation Made Simple. *Phys. Rev. Lett.* 1996, 77, 3865-3868.

(21) Kresse, G. From Ultrasoft Pseudopotentials to the Projector Augmented-Wave Method. *Phys. Rev. B* 1999, 59, 1758-1775.

(22) Grimme, S. Semiempirical GGA-type Density Functional Constructed with a Long-Range Dispersion Correction. *J. Comput. Chem.* 2006, 27, 1787-1799.

(23) Heyd, J.; Scuseria, G. E.; Ernzerhof, M. Hybrid Functionals Based on a Screened Coulomb Potential. *J. Phys. Chem. C* 2003, 118, 8207-8215.

(24) Barnett, R.; Landman, U. Born-Oppenheimer Molecular-Dynamics Simulations of Finite Systems: Structure and Dynamics of $(H_2O)_2$. *Phys. Rev. B* 1993, 48, 2081-2097.

(25) Togo, A.; Oba, F.; Tanaka, I. First-Principles Calculations of the Ferroelastic Transition between Rutile-Type and $CaCl_2$-type $SiO_2$ at High Pressures. *Phys. Rev. B.* 2008, 78, 134106.

(26) Monconduit, L.; Evain, M.; Brec, R.; Rouxel, J.; Canadell, E. *C.R. Acad. Sci. Paris* 1993, 316, 25.

(27) Gareh, J.; Boucher, F.; Evain, M. *Eur. J. Solid State Inorg. Chem.* 1996, 33, 355.

(28) Wang, W.; Dai, S.; Li, X.; Yang, J.; Srolovitz, D. J.; Zheng, Q. Measurement of the Cleavage Energy of Graphite. *Nat. Commun.* 2015, 6, 7853.

(29) Björkman, T.; Gulans, A.; Krasheninnikov, A. V.; Nieminen, R. M. Van der Waals Bonding in Layered Compounds from Advanced Density-Functional First-Principles Calculations. *Phys. Rev. Lett.* 2012, 108, 235502.

(30) Li, F.; Liu, X.; Wang, Y.; Li, Y. Germanium Monosulfide Monolayer: A Novel Two-Dimensional Semiconductor with a High Carrier Mobility. *J. Mater. Chem. C* 2016, 4, 2155-2159.

(31) Qiao, M.; Liu, J.; Wang, Y.; Li, Y.; Chen, Z. $PdSeO_3$ Monolayer: Promising Inorganic 2D Photocatalyst for Direct Overall Water Splitting Without Using Sacrificial Reagents and Cocatalysts. *J. Am. Chem. Soc.* 2018, 140, 12256-12262.

(32) Wang, Y.; Qiao, M.; Li, Y.; Chen, Z. A Two-Dimensional CaSi Monolayer with Quasi-Planar Pentacoordinate Silicon. *Nanoscale Horiz.* 2018, 3, 327-334.

(33) Peng, R.; Ma, Y.; Huang. B.; Dai, Y. Two-Dimensional Janus PtSSe for Photocatalytic Water Splitting under the Visible or Infrared Light *J. Mater. Chem. A* 2019, 7, 603-610.

(34) Cadelano, E.; Palla, P. L.; Giordano, S.; Colombo, L. Elastic Properties of Hydrogenated Graphene. *Phys. Rev. B* 2010, 82, 235414.

(35) Peng,R.; Ma, Y.; He, Z.; Huang,B.; Kou, L.; Dai, Y. Single-Layer $Ag_2S$: A Two-Dimensional Bidirectional Auxetic Semiconductor. *Nano Lett.* 2019, 192, 1227-1233.

(36) Andrew, R. C.; Mapasha, R. E.; Ukpong A. M.; Chetty, N. Mechanical Properties of Graphene and Boronitrene. *Phys. Rev. B* 2012, 85, 125428.

(37) Michel, K. H.; Verberck, B. Theory of Elastic and Piezoelectric Effects in Two-dimensional Hexagonal Boron Nitride. *Phys. Rev. B* 2009, 80, 224301.

(38) Gao, Z.; Dong, X.; Li, N.; Ren, J. Novel Two-Dimensional Silicon Dioxide with In-Plane Negative Poisson's Ratio. *Nano Lett.* 2017, 17, 772-777.

(39) Xu, X.; Ma, Y.; Huang, B.; Dai, Y. Two-Dimensional Ferroelastic Semiconductors in Single-Layer Indium Oxygen Halide InOY (Y = Cl/Br). *Phys. Chem. Chem. Phys.* 2019, 21, 7440-7446.

(40) Mills, G.; Jónsson, H.; Schenter, G. K. Reversible Work Transition State Theory: Application to Dissociative Adsorption of Hydrogen. *Surf. Sci.* 1995, 324, 305-337.

(41) Kou, L.; Ma, Y.; Tang, C.; Sun, Z.; Du, A.; Chen, C. Auxetic and Ferroelastic Borophane: A Novel 2D Material with Negative Possion's Ratio and Switchable Dirac Transport Channels. *Nano Lett.* 2016, 16, 7910-7914.

(42) Li, W.; Li, J. Ferroelasticity and Domain Physics in Two-Dimensional Transition Metal Dichalcogenide Monolayers.

*Nat. Commun.* 2016, 7, 10843-10846.

(43) Qiao, J.; Kong, X.; Hu, Z.-X.; Yang, F.; Ji, W. High-Mobility Transport Anisotropy and Linear Dichroism in Few-Layer Black Phosphorus. *Nat. Commun.* 2014, 5, 4475.

(44) Chen, J.; Xi, J.; Wang, D.; Shuai, Z. Carrier Mobility in Graphyne Should Be Even Larger than That in Graphene: A Theoretical Prediction. *J. Phys. Chem. Lett.* 2013, 4, 1443-1448.

(45) Sun, S. Meng, F.; Xu, Y.; He, J.; Ni, Y.; Wang, H. Flexible, Auxetic and Strain-Tunable Two Dimensional Penta-$X_2$C Family as Water Splitting Photocatalysts with High Carrier Mobility. *J. Mater. Chem. A* 2019, 7, 7791-7799.

(46) Li, X.; Mullen, J.; Jin, Z.; BorysenkoK, M.; Nardelli, M.; Kim, K. Intrinsic Electrical Transport Properties of Monolayer Silicene and $MoS_2$ from First Principles. *Phys. Rev. B* 2013, 87, 588-591.

(47) Fei, R.; Yang, L. Strain-Engineering the Anisotropic Electrical Conductance of Few Layer Black Phosphorus. *Nano Lett.* 2014, 14, 2884-2889.

(48) Huang, X.; Paudel, T. R.; Dong, S.; Tsymbal, E. Y. Hexagonal Rare-Earth Manganites as Promising Photovoltaics and Light Polarizers. *Phys. Rev. B* 2015, 92, 125201.